# How do chaos and turbulence affect the predictability of natural complex fluid flow systems?

Dragutin T. Mihailović[a, *], Slavica Malinović-Milićević[b], Francisco Javier Frau[c], Vijay P. Singh[d], and Jeongwoo Han[e]

[a]Department of Physics, Faculty of Sciences, University of Novi Sad, Novi Sad, Serbia; guto@df.uns.ac.rs

[b]Geographical Institute "Jovan Cvijić", Serbian Academy of Sciences and Arts, Belgrade, Serbia; s.malinovic-milicevic@gi.sanu.ac.rs

[c]Andean Regional Center, National Institue of Water, Mendoza, Argentina; fjfrau@ina.gob.ar

[d]Department of Biological and Agricultural Engineering and Zachry Department of Civil & Environmental Engineering, Texas A&M University, College Station, TX 77843-2117, USA; vsingh@tamu.edu

[e]Department of Biological and Agricultural Engineering, Texas A&M University, College Station, TX, USA; han820124@tamu.edu

*Correspondence: guto@df.uns.ac.rs; tel.: +38121458449

**Abstract**

Natural complex fluid flow systems exhibit turbulent and chaotic behavior that determines their high-level complexity. Chaos has an accurate mathematical definition, while turbulence is a property of fluid flow without an accurate mathematical definition. Using the Kolmogorov complexity (KC) and its derivatives (KC spectrum and its highest value), permutation entropy (PE), and Lyapunov exponent (LE), we considered how chaos and turbulence affect the predictability of natural complex fluid flow systems. This paper applied KC, Kolmogorov complexity spectrum, PE, and LE measures to investigate the turbulent and chaotic behaviors of the monthly streamflow of rivers from Bosnia and Herzegovina, the United States, and the Mendoza Basin (Argentina) and evaluated their time horizons using the Lyapunov time (LT). Based on the measures applied for river streamflow, we derived four modes of the interrelationship between turbulence and chaos. Finally, using those modes, we clustered rivers with similar time horizons representing their predictability. In summary, the calculated quantities of the measures were in the following intervals: (i) KC (0.484, 0.992), (ii) PE (0.632, 0.866), (iii) LE (0.108, 0.278), and (iv) LT (3.4, 9.3 months).

## 1. Introduction

*1.1 Turbulent and chaotic behavior of rivers*

This study begins with defining the concepts of complexity and chaos that were used in the paper. Complexity is a nontrivial regularity stemming from the internal structure of a system. There is no unique explanation of complexity, and the probable most general definition is that a system exhibits complexity when its behavior cannot be easily explained by examining its components (Mihailović et al., 2023a). Chaos is a mathematical and scientific concept referring to a type of behavior exhibited by certain nonlinear dynamical systems. It is characterized by extreme sensitivity to initial conditions: small changes in the starting conditions of the system can lead to significantly different consequences over time. A mathematical definition of chaos often involves three key properties: sensitivity to initial conditions, topological mixing, and dense periodic orbits. Chaotic systems are deterministic, meaning that their future behavior is entirely determined by their initial conditions and the rules that govern their dynamics. However, the long-term behavior of chaotic systems is highly unpredictable and random. This is because even tiny errors or uncertainties in the initial conditions amplify as time progresses such that trajectories are divergent and long-term predictions are impossible.

We have a strong perceptual impression that a wide river in the lowland is calm and not much turbulent. Birnir (2008) theoretically proved the existence of solutions that describe turbulent flow in rivers. The Reynolds number of rivers ($Re_{riv}$) is calculated as $Re_{riv} = \overline{D}\,\overline{V}/\nu$, where $\overline{D}$ is the average depth of flow, $\overline{V}$ is the average velocity, and $\nu$ the kinematic viscosity. Streams and/or rivers have a typically large $Re_{riv}$ ($Re_{riv} = 10^5 - 10^6$) (Dingman, 1984). Turbulence is a state of fluid flow that is characterized by irregular fluctuations in velocity, pressure, and other physical quantities having much more degrees of freedom than flows in a chaotic mode. This chaotic behavior arises due to the presence of instabilities in the fluid flow, such as vortices, eddies, and other nonlinear phenomena. These instabilities cause energy to cascade from larger scales to smaller scales, resulting in a complex and chaotic flow pattern. All chaotic flows are not necessarily turbulent, which is vividly described by (Li, 2014): "when the Reynolds number is large, violent fully developed turbulence is due to 'rough dependence on initial data' rather than chaos which is caused by 'sensitive dependence on initial data'; when the Reynolds number is moderate, turbulence is due to chaos." Conjecturing from the description by Li (2014), rivers are *par excellence* complex systems that include turbulent and chaotic behavior determining a high-level complexity of their flow. For the sake of clarity herein, we will distinguish between chaotic behavior and turbulence. Indeed, chaos has an accurate mathematical definition, while turbulence is a property of fluid flow but without an accurate *mathematical* definition.

In rivers, you can observe both spatial and temporal irregular fluctuations at different scales. These fluctuations are influenced by the hydrological regime that is characterized by nonlinearly intertwined factors among stochastic hydrometeorological forcings, physiography, channel morphology, sediment transport, and land use. Spatial irregular fluctuations are affected by the variations of the geomorphic features of a river and hydrologic regime that are regionally different. Temporal irregularities refer to variations in river properties and processes over time, including seasonal fluctuations, diurnal variations, short-therm variations, and long-term trends. Since the aforementioned spatial and temporal irregular fluctuations occur as three-dimensional eddies, it is impossible to prove whether these are stochastic or chaotically deterministic. Consequently, turbulence can be either (i) one example of the physical manifestation of deterministic chaos or (ii) a stochastic, non-chaotic manifestation of the solution to the nonlinear fluid flow problem at *high* Reynolds numbers.

The natural fluid flow systems are complex. So, to understand their behavior, the term "chaos" is used to define the situation where complex and random behaviors arising from nonlinear deterministic systems are sensitive to initial conditions (De Domenico and Ghorbani, 2010). In other words, this definition includes the three essential intrinsic properties: (1) nonlinear mutually dependent relationship; (2) unseen determinism and order; and (3) sensitivity to initial conditions. On the other hand, turbulent flow always shows a high degree of complexity, which is one of the most discriminating properties of complex systems. Complexity, by itself, cannot be modeled, so the measures using complexity as an avenue to detect chaos are entangled in a rather complicated manner (Mihailović et al., 2023b). To understand interdependence better, one can use Fig. 1 which shows a scatter plot of Kolmogorov complexity (KC) versus Lyapunov exponent (LE) for different values of the logistic parameters in the difference logistic equation $x_{n+1} = rx_n(1 - x_n)$, where $1 \leq r \leq 4$ is the logistic parameter and $0 \leq x \leq 1$, and $n$ is number of step. This equation is used in many examples since it shows an unknown and unexpected trait that is completely distant from our intuition. Figure 1 shows that a wider scatter up to KC of ~0.5, afterward that KC and LE enter into a linear trend.

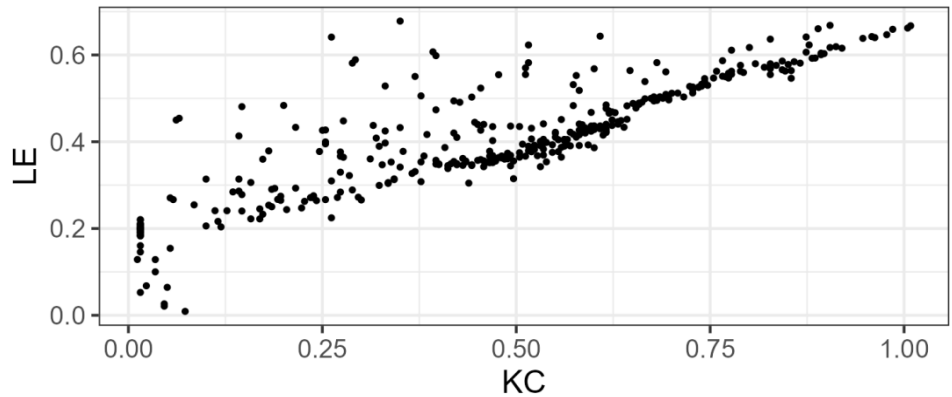

**Fig. 1.** Scatter plot of Kolmogorov complexity (KC) versus Lyapunov exponent (LE) for the logistic equation with the logistic parameter ranging between 3.6 and 3.99 changing with the step of 0.001.

It is quite natural to raise the question: how do chaos and turbulence affect the *predictability* of river flow (hereafter streamflow)? The predictability of streamflow usually refers to (i) the time evolution of the natural fluid flow system from which it is possible to obtain information and (ii) the content of that information. We will focus on models that predict the state of natural fluid flow systems for a longer period of time and a larger spatial scale. Due to the complexity of natural fluid flow systems, it is difficult to estimate their *prediction horizon* (Mihailović et al., 2019; Mihailović et al., 2023a).

*1.2 Predictability of streamflow*

In mathematics and sciences, a timescale known as the *Lyapunov time* (LT), which is called *prediction horizon*, is widely used to represent the time horizons that is a limit of the period for a dynamical system to be predictable before entering the chaotic regime with unpredictability. So, it indicates the *limits of predictability*. LT is defined as the inverse of the largest Lyapunov exponent of the time series and is expressed in the units of the observed or derived time series. Estimation of the Lyapunov time is related to computational or intrinsic uncertainties, often leading to an overestimation of the authentic value of the period. To correct this overestimation, Mihailović et al. (2019) recommended using the *Kolmogorov time* as the inverse of the Kolmogorov complexity. This time can be interpreted as the length of the time interval within which complexity remains unchanged, meaningfully reducing the size of the actual prediction horizon. River regimes are simple, mixed, or complex, and one question is how these regimes are related to turbulence, chaos, and predictability.

The prediction horizon of streamflow is a consequence of entangled hydro-meteorological forcing, physiographic, and human activities. The prediction horizon of streamflow can be better assessed by information measures (Kolmogorov complexity, Kolmogorov complexity spectrum, permutation entropy, and Lyapunov exponent in this paper) that provide more reliable information.

## 2. Method

*2.1 Kolmogorov complexity and its derivatives*

Kolmogorov complexity $K(x)$ is one of the most fundamental concepts in algorithmic information theory (Kolmogorov, 1965). The Kolmogorov complexity (KC) of an object represented by a finite binary sequence is defined as the minimum length of the binary sequence needed to reconstruct the whole object. Therefore, the object is more complex if it is described by a longer sequence and vice versa. Its limitation is that it is generally incomputable and can only be approximated. In practical applications, KC can be approximated by using some data compressor $C(x)$; furthermore, $C(x) \approx K(x)$ for most sequences (Grünwald and Vitányi, 2008). The most famous algorithm, Lempel-Ziv Algorithm (LZA) (Lempel and Ziv, 1976) was improved by Welch (1984). LZA counts the minimal number of distinct patterns in a given time series. Note that LZA, although commonly applied to longer sequences in many sciences, is inapplicable to short time series (less than 50 samples). Although LZA can be found in many papers, we will describe it briefly herein to facilitate understanding. LZA for calculating the complexity of a time series $X(x_1, x_2, x_3, \ldots, x_N)$ includes the following steps: (1) Creating a binary time series $s(i), i = 1, 2, \ldots, N$ with 0 and 1, according to the rule of $s(i) = 0$ if $x_i < x_t$ or 1 if $x_i > x_t$, where $x_t$ is a threshold and $i$ is a time instance. The threshold is commonly selected as the mean value of the time series, but other encoding schemes are also available; (2) calculating the complexity counter $c(N)$ that is defined as the minimum number of distinct patterns contained in the given binary time series. This counter is a function of the length of sequence $N$. The value of $e(N)$ approaches an ultimate value $c(N)$ as $N$ approaches infinity, i.e., $c(N) = O(b(N))$ and $b(N) = log_2 N$; and (3) Calculating a normalized information measure $C_k(N)$, which is defined as $C_k(N) = c(N)/b(N) = c(N)/log_2 N$. For a nonlinear time series, $C_k(N)$ varies between 0 and 1, although it can be larger than 1.

The KC of time series (i) cannot distinguish between time series with different amplitude variations and those with similar random components; and (ii) after the binarization of a time series, its complexity can be lost because of the applied procedure. In the complexity analysis of time series, two measures are used: (i) Kolmogorov complexity spectrum (KC spectrum) and (ii) the highest value of KC spectrum (KCM), introduced by (Mihailović et al., 2015) who described the procedure for calculating the KC spectrum. The KC spectrum $C(c_1, c_2, c_3, \ldots, c_N)$ is derived from the time series $X(x_1, x_2, x_3, \ldots, x_N)$ in the following way: Each sample in $X(x_1, x_2, x_3, \ldots, x_N)$ is used as a threshold and then the corresponding KC is calculated forming the time series $C(c_1, c_2, c_3, \ldots, c_N)$. Thus, we obtain the KC spectrum (KC versus amplitude). The flow chart for its calculation can be found in Mihailović et al. (2019). This spectrum is suitable for the analysis of the range of amplitudes in a time series that represents a complex system with highly enhanced stochastic components. It was used in the analysis of processes related to natural fluid flow (Lade et al., 2019; Sharma et al., 2018). The highest value of the KC complexity spectrum $K_m^C$, which can be referred to as KCM, is written as $K_m^C = max\{c_i\}$.

## 2.2 Calculation of permutation entropy

As a statistical measure, permutation entropy describes the complexity of a time series through phase space reconstruction (Bandt and Pompe, 2002). It takes into account the non-linear behavior of the time series. The advantages of this measure are (i) applicability to real data, (ii) robustness if observational noise is present, and (iii) invariance to non-linear transformations. For $N$ sample time series $\{x(i): 1 \leq i \leq N\}$, all permutations of $\pi$ of order $m$ ($m < N$), amounted to $m!$, are considered. The relative frequency for each permutation $\pi$ is

$$p(\pi) = \frac{\#\{i | 0 \leq i \leq N-m, (x_{i+1},...,x_{i+m}) \text{ is of type } \pi\}}{N-m+1}. \tag{1}$$

When the underlying stochastic process satisfies a very weak stationary condition that $x_i < x_{i+k}$ for $k \leq m$ is independent of $i$, the relative frequency $p(\pi)$ converges to the exact probability if $N \to \infty$.

The permutation entropy of order $m \geq 2$ is defined as $H(m) = \sum_{i=1}^{m!} p(\pi_i) \log p(\pi_i)$. The value of $H(m)$ is always $0 \leq H(m) \leq \log(m!)$, where the lower bound is attained for monotone time series (increasing or decreasing), and the upper bound is attained for an identically independent random sequence when all possible permutations have the same probability. For chaotic time series, $H(m)$ increases almost linearly with $m$.

## 2.3 Calculating the Lyapunov exponent for time series

River flow processes, like many others in nature, can be characterized by their sensitivity to initial conditions. If $\delta(t)$ is the distance between two nearby orbits in the phase space at some time $t$, the evolution of the sensitivity to initial conditions $\varphi(t) = \delta(t)/\delta(0)$ can show quite different behaviors for deterministic as well as stochastic dynamical systems as (Gao et al., 2007).

$$\frac{d\varphi(t)_{Det.(t)}}{dt} = \lambda_{max}\varphi(t)_{Det.(t)} \qquad (2)$$

$$\frac{d\varphi(t)_{Stoc.(t)}}{dt} = t^H \qquad (3)$$

where $\lambda_{max}$ is called the largest Lyapunov exponent and $H$ is called the Hurst exponent which is a statistical measure that is used to study scaling properties in time series. For a more detailed discussion, reference has been made to Qian and Rasheed (2004).

Positive Lyapunov exponent (LE) indicates that small fluctuations can lead to drastically different system behavior (small differences in the initial state lead to large differences in a later state). Because the rate of separation can be different for different orientations of the initial separation vector, the largest value of a spectrum of Lyapunov exponents is commonly used to be LE. A positive value of this exponent is taken as an indicator denoting that a dynamical system is chaotic. In this study, we obtained LE for the standardized monthly streamflow time series by applying the Rosenstein algorithm (Rosenstein et al., 1993), which was implemented in the MATLAB program (Shapour, 2009). Let us note that this measure has one drawback. If the embedding theory is used to build chaotic attractors in the reconstruction space, then additional "spurious" Lyapunov exponents may appear.

*2.4 Estimation of prediction horizon*

We already mentioned that the prediction time (LT) is defined as the inverse of the largest Lyapunov exponent (LLE) of the considered time series, i.e., LT=1/LLE. The information content of forecasts of approximately stationary quantities tends to decline as the prediction horizon increases. Therefore, there exists a maximum prediction horizon beyond which forecasts cannot provide detectably more information about the forecasting variable (Galbraith and Tkacz, 2007). However, still there isn't a consensus about the relationship between the maximum prediction horizon and LLE in time series prediction. Thus, in some papers, LT was calculated as 1/ LLE, while some others argued that LT should be multiplied by some values (De Domenico and Ghorbani, 2010; Gao et al., 2007). 1/LLE gives the correct timescale for the prediction horizon. According to Small (2014), however, whether using LT or LT multiplied by some constant is right depends on the (un-)certainty with which we wish to be able to make predictions. In this paper, we employed the estimation of the prediction horizon as 1/LLE.

## 3.  Description of data

*3.1. Criteria for the selection of time series*

This paper aims to explore the connection between chaos and complexity and its influence on the time horizon for streamflow time series of rivers with different attributes. For this purpose, we chose: (a) 11 gauge stations from seven rivers in Bosnia and Herzegovina; (b) 5 gauge stations from the Mendoza Basin of the Andes region in Argentina; and (c) 1879 gauge stations from different watersheds in the U.S. The streamflow data obtained for the study encompasses diverse hydroclimate and physiography so that they express probably all combinations of the relationship between complexity and chaos. The gauge stations were classified into three types based on the elevation as follows (Meybeck et al., 2001): (i) lowlands, when the mean elevation of a watershed is 0-200 m, hereafter, it is referred to as L type, (ii) platforms and hills, when the mean elevation of a watershed is 200-500 m, hereafter, it is referred to as H type, and (iii) mountains, when the mean elevation of a watershed is 500-6000 m, hereafter, it is referred to as M type.

*3.2. Site and data description*

*Bosnia and Herzegovina*. The territory of Bosnia and Herzegovina is positioned in the western Balkan surrounded by Croatia to the north and southwest, Serbia to the east, and Montenegro to the southeast. It lies between latitudes 42º and 46º N and longitudes 15º and 20º E. The country is mostly mountainous, encompassing the central Dinaric Alps. The northeastern parts reach into the Pannonian basin, while it borders the Adriatic Sea in the south. Dinaric Alps generally run in the east-west direction and get higher towards the south. The gauge stations were located either down-, upstream, or both in a watershed (Fig. 2). Their characteristics are given in Table 1, while time series are shown in Fig. 3.

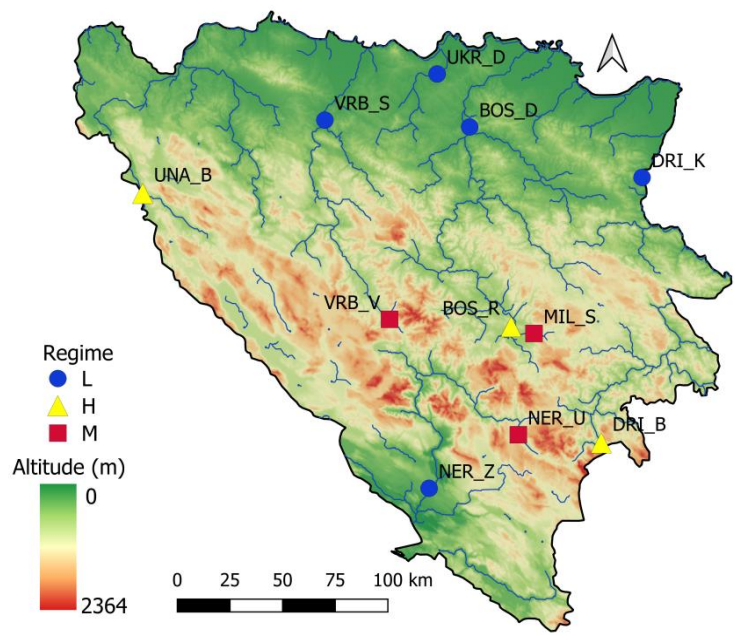

**Fig. 2.** Elevation of Bosnia and Herzegovina with the locations of eleven gauge stations for seven rivers used in the study. The abbreviations for rivers are listed in Table 1, while geometric shapes indicate the river regimes.

**Table 1**. Rivers in Bosnia and Herzegovina used in the study with the corresponding monthly streamflows (FR) for the period 1965-1986 and their regime classification (Meybeck et al., 2001) as lowland (alt < 200 m; L regime), highland (200< alt < 500 m; H regime), and mountains (500< alt < 6000 m; M regime).

| Rivers and gauge-stations | Abb. | Long. (°E) | Lat. (°N) | Alt. (m) | FR ($m^3/s$) | Regime | Period |
|---|---|---|---|---|---|---|---|
| River Neretva to Zitomislić | NER_Z | 17°47' | 43°12' | 16 | 252.0 | L | 1965-1986 |
| River Neretva to Ulog | NER_U | 18°19' | 43°25' | 641 | 8.0 | M | 1965-1986 |
| River Bosna to Doboj | BOS_D | 18°06' | 44°45' | 137 | 172.0 | L | 1965-1986 |
| River Bosna to Reljevo | BOS_R | 18°20' | 43°53' | 478 | 29.0 | H | 1965-1986 |
| River Drina to Kozluk | DRI_K | 19°07' | 44°30' | 121 | 380.0 | L | 1965-1986 |
| River Drina to Bastasi | DRI_B | 18°48' | 43°22' | 425 | 155.0 | H | 1965-1986 |
| River Miljacka to Sarajevo | MIL_S | 18°26' | 43°51' | 539 | 5.0 | M | 1965-1986 |
| River Una to Martin Brod | UNA_B | 16°08' | 44°30' | 310 | 54.0 | H | 1965-1986 |
| River Ukrina to Derventa | UKR_D | 17°55' | 44°59' | 104 | 17.0 | L | 1965-1986 |
| River Vrbas to D. Selo | VRB_S | 17°14' | 44°48' | 144 | 111.0 | L | 1965-1968 |
| River Vrbas to G. Vakuf | VRB_V | 17°35' | 43°56' | 647 | 3.7 | M | 1965-1986 |

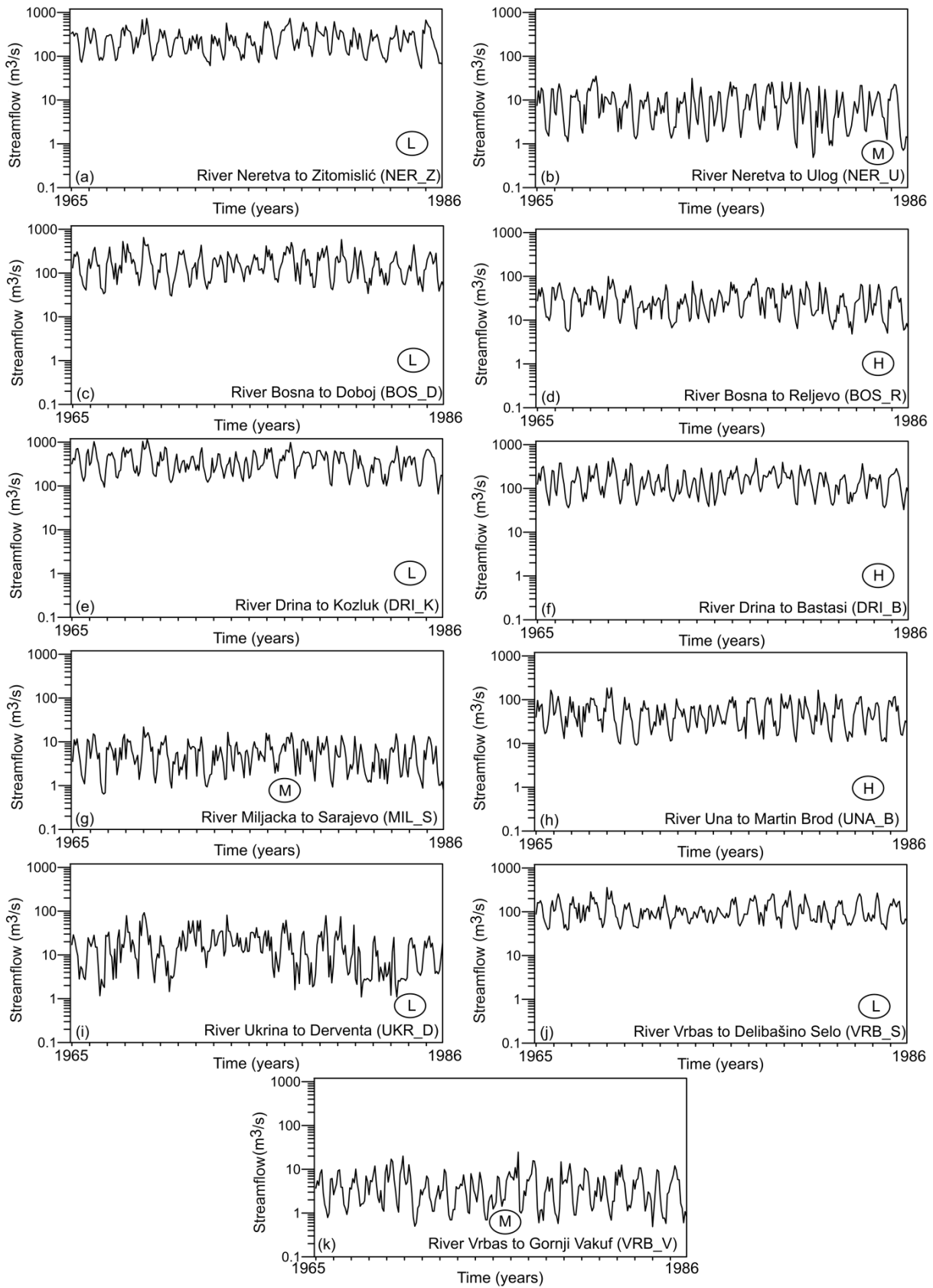

**Fig. 3.** Monthly streamflow time series at 11 gauge stations for seven rivers in Bosnia and Herzegovina for the period 1965–1986.

*The Mendoza Basin in Argentina.* Monthly streamflow data was obtained from the National Hydrological Network (RHN) from the Secretary of Infrastructure and Water Policy. The basin is located between latitudes 31º and 33ºS in the Central Andean Region, to the west of Argentina (Fig. 4). The Mendoza River basin has three upstream tributaries - Vacas, Cuevas, and Tupungato Rivers - that confluence with the Mendoza River. The gauge stations for the three upstream tributaries are located at the outlet of each sub-watershed. Two gauge stations are located on the mainstream of Mendoza River before the Potrerillos dam. Cacheuta gauge station near the outlet of the Mendoza River basin is currently unavailable since it was dismounted with the dam construction. Their characteristics are given in Table 2, while time series are shown in Fig. 5.

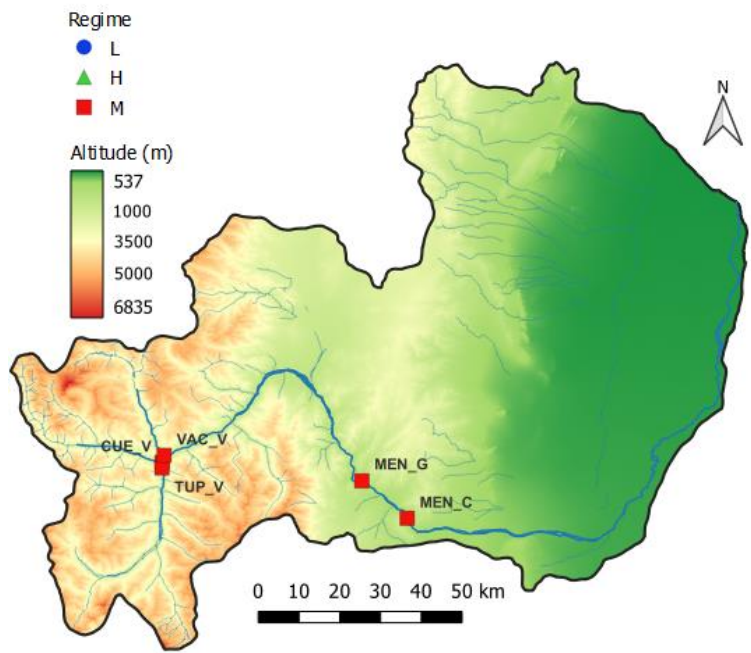

**Fig. 4.** Elevation of Mendoza basin (Argentina) with the locations of five gauge stations for four rivers used in the study. The abbreviations for rivers are listed in Table 2.

**Table 2**. Rivers in Mendoza Basin (Argentina) used in the study with the corresponding flow rates (FR – mean for the period indicated) and their regime classified as did Table 1.

| Rivers and gauge-stations | Abb. | Long. (°W) | Lat. (°S) | Alt. (m) | FR (m$^3$/s) | Regime | Period |
|---|---|---|---|---|---|---|---|
| River Tupungato to Punta de Vacas | TUP_V | 69°46' | 32°53' | 2462 | 22.4 | M | 1954-2022 |
| River Cuevas to Punta de Vacas | CUE_V | 69°46' | 32°52' | 2406 | 6.4 | M | 1955-2022 |
| River Vacas to Punta de Vacas | VAC_V | 69°46' | 32°51' | 2400 | 4.2 | M | 1954-2022 |
| River Mendoza to Guido | MEN_G | 69°14' | 32°55' | 1408 | 44.0 | M | 1956-2022 |
| River Mendoza to Cacheuta | MEN_C | 69°07' | 33°01' | 1250 | 50.2 | M | 1909-1990 |

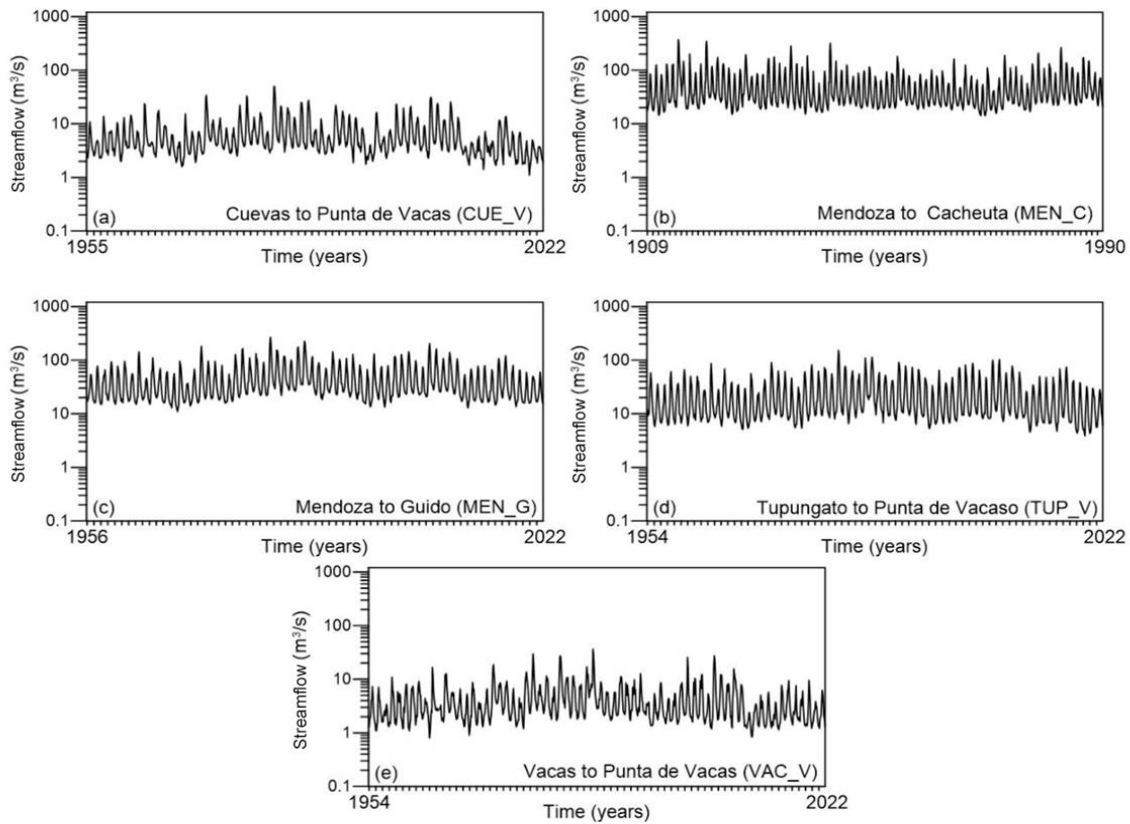

**Fig. 5.** Monthly streamflow time series at five-gauge stations for four rivers in Mendoza for the periods indicated in Table 2.

*Rivers in the United States*. Monthly naturalized streamflow data for a period from 1950 to 2015 was obtained from the U.S. Geological Survey (USGS) Science Base Catalog. The naturalized streamflow is a simulated data for 2,622,273 stream reaches, which are defined by National Hydrography Dataset (NHD) Version 2.0, across the continental U.S. using the random forest ensemble (Miller et al., 2018). We used gauge stations at the outlet of each hydrologic unit code 8 (HUC8) watershed. Using 2,622,273 flow from all reach segments provides redundant information, and using the mainstream of the midsize watershed can convey enough of the information inherited by its tributaries. Besides, since the model was calibrated at the gauge point locations, the estimated naturalized flow is accurate at the gauge stations that exist. Therefore, this study used naturalized flow at 1879 gauge stations (Fig. 6). Although the gauge stations are located in the outlet of watersheds, they also can be classified into three different regimes as shown in Fig. 6. The flow rate and altitude were averaged in the classified regimes (i.e., L, H, and M regimes) and are given in Table 3, while the ensemble averages of time series in the classified regimes are shown in Fig. 7.

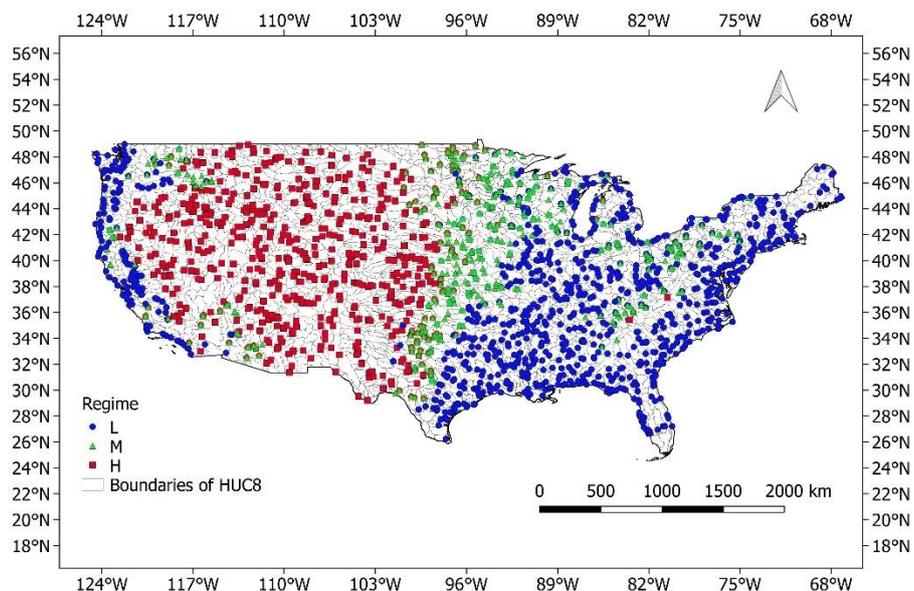

**Fig. 6.** Spatial distributions of gauge stations of 1879 U.S. rivers for the period 1950-2015. Stations are classified into regimes according to the criterion in Table 1.

**Table 3**. Rivers in the United States used in the study with the corresponding monthly streamflows (FR) and their classified regimes as did Table 1.

| Regime | Mean altitude (m) | Mean FR (m$^3$/s) | Period |
|---|---|---|---|
| L | 76 | 253.5 | 1950-2015 |
| H | 310 | 71.2 | 1950-2015 |
| M | 1172 | 31.2 | 1950-2015 |

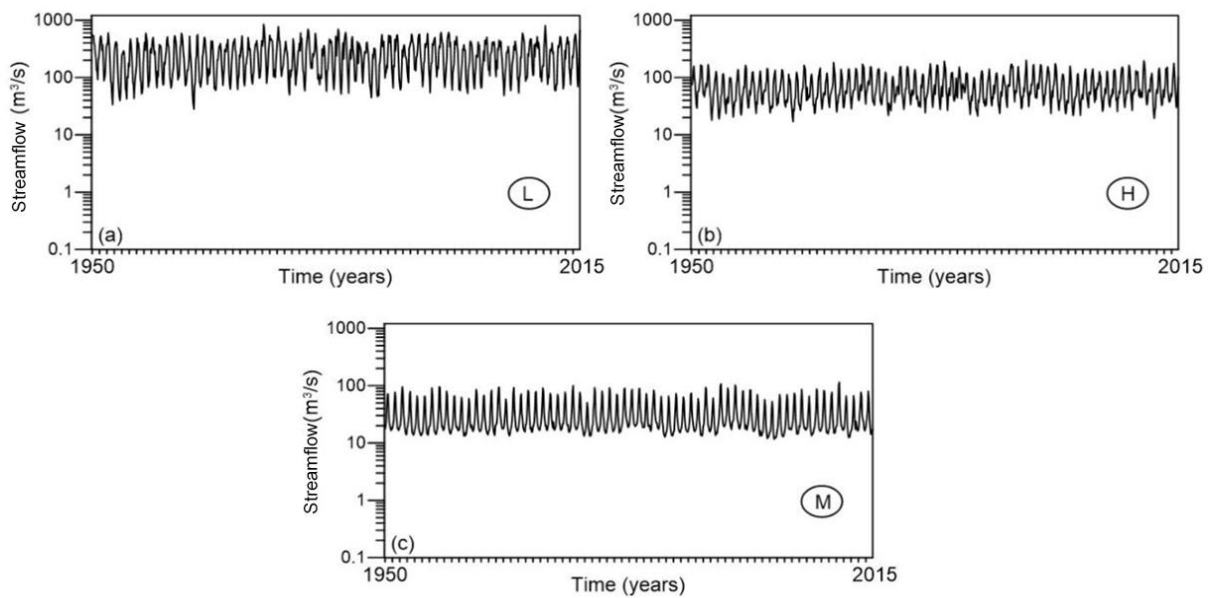

**Fig. 7.** Monthly streamflow time series at 1879 gauge stations for rivers in the U.S. averaged over all regimes: (a) L (912), H (408), and M (559) for the periods indicated in Table 3. The number of gauge stations is in the brackets.

## 4. Results and discussion

*4.1 Pathway of analysis*

Physics permeates today's world so much that we often cannot even see it, but even if we see it, it happens that we cannot interpret it clearly; that is noted as "the entangled dance of physics" by (Benka, 2006). In the following analysis, we will encounter one such situation in which there is a sort of synergy between the two properties of complex systems - chaos and complexity having the greatest impact on streamflow predictability. Although there is still a dilemma whether chaos is an actual state or just a name for rules we have not discovered yet, it is certain that chaos is one of the properties of complex systems. On the other hand, complexity is also a property of a complex system, and it is one of the most discriminating properties that cannot be modeled. Finally, let us mention the relationship between chaos and complexity with an illustrative comment by Bertuglia and Vaio (2005): "Stand-alone chaos and complexity have absolutely nothing to do with generating formal function. Chaotic systems are not necessarily complex, and complex systems are not necessarily chaotic." Our intention is to use the information measures KC and PE to get closer to the explanation of the form of the synergy of chaos and complexity in natural fluids as complex systems.

*4.2 Outcomes of information measures for Bosnia and Herzegovina and Mendoza River Basin time series*

The values of the calculated information measures KC, PE, and LE of time series for the rivers in Bosnia and Herzegovina (BiH) and Mendoza Basin (MB) are given in Table 4. The aforementioned measures were also calculated for rivers in the U.S., but they are discussed through their mean values for all three regimes.

**Table 4.** Kolmogorov complexities (KC), permutation entropy (PE), and Lyapunov exponent (LE) values of the monthly streamflow time series for rivers in Bosnia and Herzegovina (BiH) and (b) rivers in the Mendoza Basin (MB).

(a) Rivers in Bosnia and Herzegovina

| Rivers and gauge-stations | Regime | Abb. | KC | PE | LE |
|---|---|---|---|---|---|

| Rivers and gauge-stations | Regime | Abb. | KC | PE | LE |
|---|---|---|---|---|---|
| River Neretva to Zitomislić | L | NER_Z | 0.918 | 0.821 | 0.186 |
| River Neretva to Ulog | M | NER_U | 1.013 | 0.839 | 0.196 |
| River Bosna to Doboj | L | BOS_D | 0.791 | 0.866 | 0.238 |
| River Bosna to Reljevo | H | BOS_R | 0.948 | 0.858 | 0.271 |
| River Drina to Kozluk | L | DRI_K | 0.823 | 0.835 | 0.327 |
| River Drina to Bastasi | H | DRI_B | 0.948 | 0.789 | 0.319 |
| River Miljacka to Sarajevo | M | MIL_S | 1.076 | 0.891 | 0.205 |
| River Una to Martin Brod | H | UNA_B | 0.948 | 0.869 | 0.243 |
| River Ukrina to Derventa | L | UKR_D | 0.981 | 0.918 | 0.312 |
| River Vrbas to Delibasino Selo | L | VRB_S | 0.918 | 0.881 | 0.244 |
| River Vrbas to Gornji Vakuf | M | VRB_V | 0.886 | 0.821 | 0.213 |

(b) Rivers in the Mendoza Basin

| Rivers and gauge-stations | Regime | Abb. | KC | PE | LE |
|---|---|---|---|---|---|
| River Tupungato to Punta de Vacas | M | TUP_V | 0.427 | 0.593 | 0.127 |
| River Cuevas to Punta de Vacas | M | CUE_V | 0.504 | 0.658 | 0.113 |
| River Vacas to Punta de Vacas | M | VAC_V | 0.557 | 0.728 | 0.122 |
| River Mendoza to Guido | M | MEN_G | 0.474 | 0.594 | 0.089 |
| River Mendoza to Cacheuta | M | TUP_V | 0.459 | 0.587 | 0.087 |

*4.3 Complexity and chaos in river flow*

The concept of the Lyapunov exponent is characteristically used in analyzing deterministic systems and may not be useful for purely stochastic systems, such as a streamflow process that is determined by a random process. However, in a purely stochastic system, the behavior of streamflow is not governed by any known deterministic dynamics. Consequently, the concept of the LE may not be applicable. If streamflow is influenced by both deterministic and stochastic factors, LE may still be a useful tool to characterize the sensitivity of the system to initial conditions. In those cases, the positive value of LE would be a good indicator that the streamflow is chaotic and sensitive to initial conditions, but has a component of stochastic behavior.

*4.3.1 Permutation entropy*

There is an opinion that PE is becoming one of the most successful complexity measures in recent years due to its simplicity, robustness, and ability to capture the essential dynamics of the measured time series. Although, in the last two decades, PE has already been applied in many scientific fields, this measure has two drawbacks: (i) the ordinal ambiguity of equal values in subsequences (Zunino et al., 2009), and (ii) the lack of information related to the sample differences in amplitude (Fadlallah et al., 2013). Parameters, such as data length $w$, embedded dimension $m$, and time delay $\tau$, can affect the PE calculation. Bandt and Pompe (2002) recommended selecting embedded dimension $m$ from the interval (3, 7). Staniek and Lehnertz (2007) established theoretically that data length $w$ between 128 and 256 data points could be considered sufficient for achieving stable and consistent PE values. In this study, we used the following values $m=5$ and $\tau=1$ as it was done in complexity analysis of turbulent river flow (Mihailović et al., 2014). In this study, the data length $w$ of the all-time series satisfied the above condition.

**Table 5.** Mean KC, PE, and LE values of the monthly streamflow in L, H, and M regimes for (a) eleven gauge stations on seven rivers in Bosnia and Herzegovina, (b) five gauge stations

on four rivers of Mendoza Basin in Argentina, and (c) 1879 gauge stations on rivers in the U.S. Lengths of time series are given in Tables 1-3.

(a) Rivers in Bosnia and Herzegovina

| Regime | Mean altitude (m) | Mean FR (m$^3$/s) | KC | PE | LE |
|---|---|---|---|---|---|
| L | 104 | 186.3 | 0.886 | 0.864 | 0.261 |
| H | 404 | 79.0 | 0.948 | 0.832 | 0.278 |
| M | 609 | 5.6 | 0.992 | 0.850 | 0.205 |

(b) Rivers in the Mendoza Basin

| Regime | Mean altitude (m) | Mean FR (m$^3$/s) | KC | PE | LE |
|---|---|---|---|---|---|
| M | 1985 | 22.4 | 0.484 | 0.632 | 0.108 |

(c) Rivers in the U.S.

| Regime | Mean altitude (m) | Mean FR (m$^3$/s) | KC | PE | LE |
|---|---|---|---|---|---|
| L | 76 | 253.5 | 0.741 | 0.860 | 0.148 |
| H | 310 | 71.2 | 0.696 | 0.857 | 0.142 |
| M | 1172 | 31.2 | 0.520 | 0.792 | 0.119 |

Inspection of Table 4 indicates that the mean PE value of MB rivers (0.632) was lower than BiH had (0.851), while LE is more than twice as small (0.108 in comparison with 0.250). In other words, the monthly streamflows of BiH rivers have higher complexity and more turbulent flow. PE is a measure of the complexity of the ordinal pattern of a time series based on the idea that complex systems tend to produce a large number of distinct ordinal patterns. A high PE indicates that the time series exhibits many different ordinal patterns, suggesting a high level of complexity, but these patterns can be generated by relatively simple computational processes. A lower number of ordinal patterns in time series typically indicates a higher degree of predictability in the considered system that generated the data. In addition, a lower LE can indicate the presence of a deterministic component of river flow. From Fig. 8a, it is seen that the U.S. rivers dominantly have high PE, with LE having a very dense concentration of points between 0.10 and 0.20. Additionally, values of mean PE (framed squares) indicate that MB rivers have a higher level of predictability than the BiH rivers (high complexity as well for LE values), while that level for the U.S. rivers is nearly between these two.

There exists an increasing trend of mean complexity of monthly streamflow of BiH rivers as the regime change from L through to M type (Table 5a). This is more obvious with KC than with PE since the H regime is a transitional regime in a relatively narrow altitude band (between 200 m and 500 m) where a large number of distinct ordinal patterns can change significantly. It somehow corresponds with our intuition, i.e., that with the increase in altitude, the number of factors that influence the complexity of streamflow also increases, thus reducing the predictability. Such a trend of PE and KC can also be observed in the streamflow of rivers in the Mendoza Basin (MB). The mean values for KC and PE for all rivers are 0.484 and 0.632, respectively (Table 5b). However, despite the fact that all rivers belong to the M regime, Table 2 shows that the difference in their altitudes is large. The mean altitude for the first three stations is 2423 m and for the other two 1329 m (Table 2). At higher altitudes, the mean of PE is 0.660, while at lower altitudes, that is 0.591. In contrast to the case of BiH and MB, the mean values of complexity (PE and KC) calculated for all regimes over 1879 U.S. rivers (Mihailović et al., 2023b) have an opposite trend (Table 5c). The downward trend of complexity, as the regime change from the M to L type, corresponds more to intuition which tells us that predictability increases going from mountain to lowland rivers. It seems that is not the case after all. This impression does not necessarily mean that this trend indicates an increase in the predictability of rivers as average altitude decreases. If only PE is considered then it may be correct but the use of the other two measures, KC and LE, can reduce the validity of this statement.

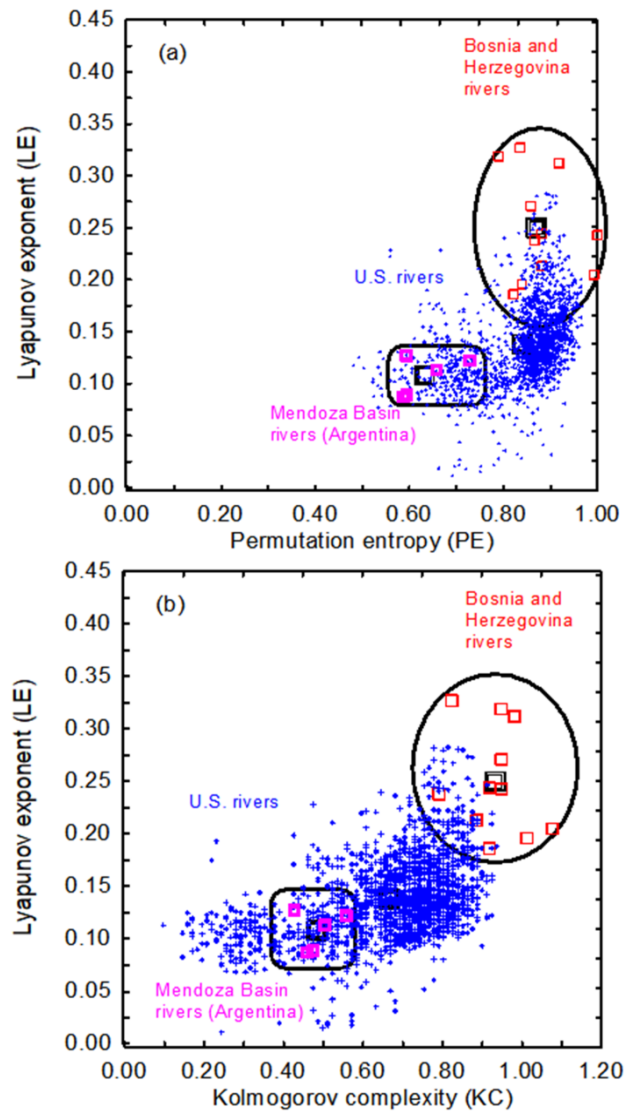

**Fig. 8.** Scatter plots of (a) PE against LE and (b) KC against LE of the monthly streamflow for BiH, MB, and the U.S. rivers. Framed squares indicate the mean for three data sets.

*4.3.2 Kolmogorov complexity*

For further analysis, it is necessary to distinguish between chaos and complexity as properties of a complex system that rivers have. Complex systems may be defined as those that have many degrees of freedom and nonlinear interactions, what is satisfied by all phenomena in nature. Therefore, they are not sufficient assumptions for analyzing complex systems, including rivers. For example, it is accepted that the edge of chaos is a transition space between order and disorder existing within numerous complex systems and consequently for rivers (Stosic et al., 2018). However, Stephens (2015) noticed that the edge of chaos cannot differentiate between complex and non-complex forms. Accordingly, chaos is a characteristic of complex systems but not a defining one. On the other hand, complexity is one of the most discriminating properties of complex systems that *cannot* be modeled. It is generally true that the complexity of a system cannot be modeled exactly, due to various reasons such as the boundaries of our knowledge, and computational power. That is why one resorts to "modeling" the complexity with more or less sophisticated models.

KC is perhaps the most useful measure for estimating the complexity of time series resulting from the evolution of complex systems. A time series have a large PE value and lower KC value at a time (Fig. 8b). It means that a high degree of complexity in terms of its ordinal patterns may be explained by a relatively small amount of information or computational complexity. Table 5 indicates that KC values are high for all regimes and all considered rivers have an increasing trend of KC as the regime change from the L to M type, similar as did PE. Comparison of KC for BiH rivers with the mean value of KC for the U.S. rivers shows the following: (i) BiH has a mean KC of 0.932 while the lowest value of 1.076 and the highest value of 0.791 were estimated from MIL_S belonging to M type and BOS_D belonging to L type, respectively and (ii) the U.S. rivers have a mean KC of 0.584, while they have the highest and lowest values of 0.936 and 0.097, respectively. The increase in KC from the L to M regime is physically explainable. Namely, the presence of instabilities in the river flow, such as vortices, eddies of different amplitudes, and other nonlinear phenomena, cause energy to cascade from larger scales to smaller scales resulting in a complex and chaotic flow pattern. Let us note that in the analysis of streamflow complexity in hydrology, the use of entropies is quite widespread (Ma et al., 2020; Wang et al., 2020).

*4.3.3 Kolmogorov complexity spectrum*

Figure 9 evidently shows the range of streamflow amplitudes for all three datasets and different regimes across the spectra of Kolmogorov complexity. The Kolmogorov spectrum can decipher the patterns in complexity of river flow that is not seeming at a single level of resolution.

From Fig. 9a, it is seen that BiH rivers that have the maximal values of complexity are very close in all three regimes. The spectra of the M regime, on the one side, and the overlapped H and L regimes, on the other side, are clearly separated. The H and L regimes have spectra overlapping in the interval between 14 $m^3s^{-1}$ and 120 $m^3s^{-1}$. MB rivers, which all belong to the M regime, have two groups of spectra that can be distinguished (Fig. 9b). The first group consists of low-streamflow amplitude spectra (up to 15 $m^3s^{-1}$) at gauge stations CUE_V and VAC_V, which are located at an average altitude of 2,423 m, while gauge station TUP_V is at the same average altitude that includes streamflow amplitudes of over 100 $m^3s^{-1}$. The second group consists of gauge stations MEN_G and MEN_C (located at the average altitude of 1,329 m) have a slightly lower maximal KC than the spectra in the first group. They cover a streamflow amplitude interval of up to 130 $m^3s^{-1}$. In the U.S. rivers (Fig. 9c) the spectra of all three regimes are located in the amplitude that ranges from 10 to 1,000 $m^3s^{-1}$. All the spectra are separated noticeably into each regime. KC decreases as the regime change from the M to L type, while the H regime forms one transition interval. Rivers belonging to the L type are typically large American rivers with lower KC. They may have a smooth, regular flow without many unexpected changes or fluctuations like (i) the lower Mississippi River having a relatively stable flow pattern due to its vast size and relatively moderate slope; and (ii) the Rio Grande River, which has a relatively predictable flow pattern due to its arid climate and relatively stable topography (Mihailović et al., 2019). 280 rivers, which amount to 14.9% of the U.S. rivers, belong to this regime (see Fig. 4 in Mihailović et al., 2023b).

Looking only at the KC in Table 5a, it is possible to conclude that the predictability for the M regime is the smallest since it has the highest KC. That may not be true since the value of LE increases when going from the M to L regime. It seems a bit confusing. However, this can address that, in a river system, a high level of complexity in the flow pattern does not necessarily point toward high chaos or a high LE. Actually, it is possible for a river to have a complex flow pattern but a lower LE (hereafter, this case is referred to as *Kl* mode having a high KC and lower LE). In this mode, the flow may appear complex due to factors such as channel morphology or the presence of obstacles, but the underlying dynamics of the flow may be relatively simple and predictable. In Fig. 8b, it is noticeable that a huge number of rivers in the U.S. belong to this regime, i.e., 955 gauge stations, which amount to 50.8% of the total number of stations, belong to the *Kl* mode.

The MB rivers that flow in the high mountain Andean region have lower KC as well as LE. This mode is called the *kl* mode. It can occur when the river flow is dominated by laminar flow, which is a smooth and ordered flow pattern with low levels of turbulence and chaos. Laminar flow can occur in some sections of a river with low velocity and smooth channel geometry, or in the boundary layer of the flow where the fluid velocity is very low. Note that, in rivers, laminar flow seldom occurs. For example, in the U. S., the following rivers are in this mode: Buffalo National River (Arkansas), Current River (Missouri),

Suwannee River (Florida), and Snake River (Wyoming and Idaho). Also, this mode encompasses the mountain river Miljacka (MIL_S), which is, in some sections, intensively channelized after the Second World War (Mihailović et al., 2014). It is important to note that these examples are not exclusively characterized by laminar flow, and the flow conditions in rivers can vary based on factors such as season, weather, and river geometry. Laminar flow is more commonly associated with smaller streams or specific sections within larger rivers.

All BiH rivers have large LE and KC (hereafter, *KL* mode). Rivers in this mode have high KC and high LE and, consequently, have more turbulent and unpredictable flow with frequent changes in water level, flow direction, and velocity. For example, the Colorado River (the U.S.) and some segments of Miljacka River (BiH) in the data set this study used are in this mode. Thus, the rivers belonging to this mode are unpredictable. However, it is worth noting that these rivers can have varying degrees of complexity and LE, and their behavior can also be affected by a range of factors such as climate change, land use practices, and different human activities. 619 (32.9% of) rivers in the U.S. are in this mode (Mihailović et al., 2023b), while all rivers in Bosnia and Herzegovina belong to the *KL* mode.

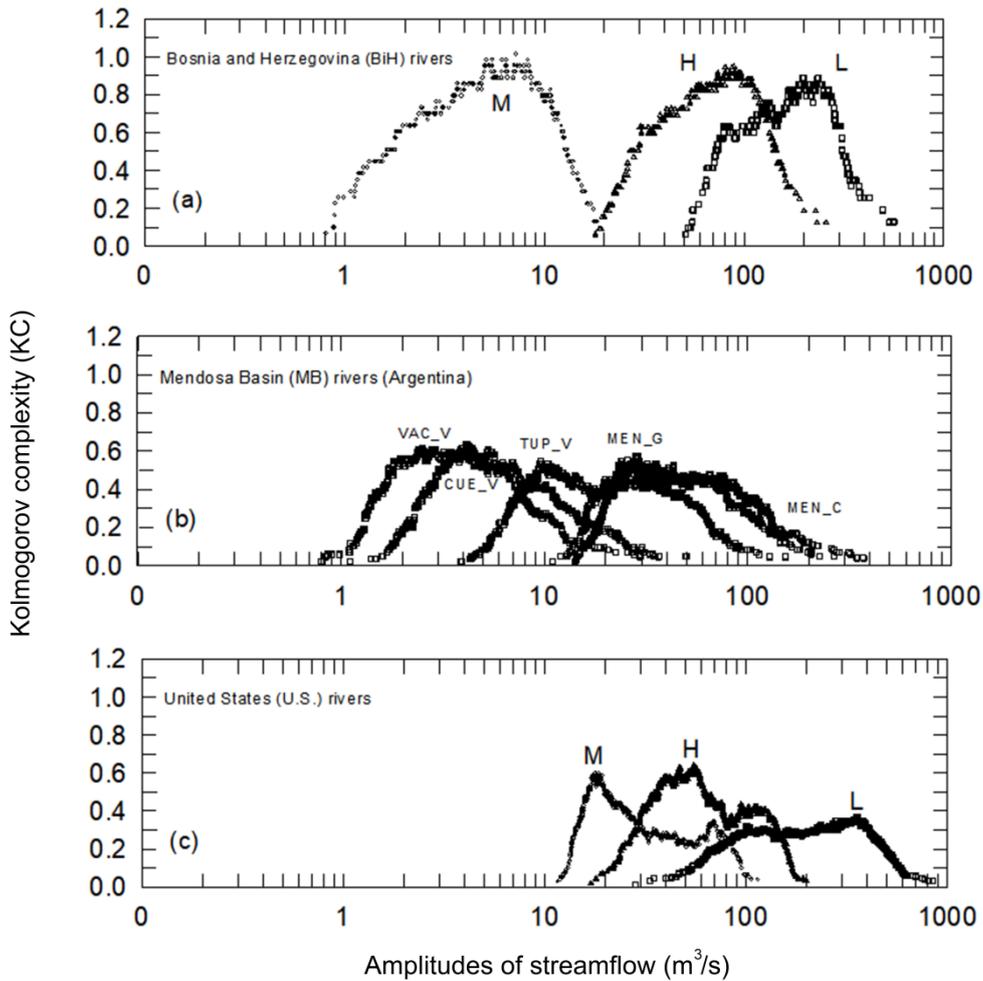

**Fig. 9.** Spectra of Kolmogorov complexity for M, H, and L regimes for rivers in BiH (a), MB (b), and the U.S (c). Abbreviations are given in Table 2.

Finally, let us consider the *kL* mode (i.e., lower KC and higher LE). A low complexity in a river flow indicates a relatively simple flow pattern with lower variations and irregularities in the flow occurring in a straight section of the river or in a section with a very smooth channel geometry. When a river has a low complexity but a high LE, it suggests that the flow is highly sensitive to small perturbations, despite the relatively simple flow pattern. These conditions can occur in the case of river flow in a transitional regime, i.e., where it is neither fully laminar nor turbulent. Ultimately, small differences in initial conditions can result in large differences in the long-term behavior of the river system. In the U.S. data set, there are 13 rivers (four rivers in the M regime, four rivers in the H regime, and five rivers in the L regime) having characteristics of *kL* mode (0.1 < KC < 0.5 and 0.17 < LE < 0.25). On average, they have KC = 0.429, LE = 0.188, and LT=5.4.

*4.3.4 Predictability and prediction horizon*

The predictability of a phenomenon, streamflow herein, usually refers to 1) the time evolution of the system from which we can obtain information and 2) the content of obtained information. While theory (i.e., a model) describes phenomena, for example, in hydrology, the concept of the model is based on interpretation in mathematics (Mihailović et al., 2023a). Scientists often apply a heuristic technique which is any approach to problem-solving via a practical method, which does not guarantee to be optimal or perfect but does satisfy either the instantaneous goals or grasp a better approach. Basically, all models in science are heuristic, having more or less good performances. In this paper, we do not deal with this aspect of predictability which is rather a technical one. Our intention is to estimate a period after which streamflow, as a dynamical system, becomes unpredictable and enters a chaotic state, indicating the limits of predictability. In other words, we search a time scale for streamflow called the prediction horizon—the Lyapunov time LT= 1/LE (expressed in the units of recorded series), where LE is the largest positive Lyapunov exponent. If LT increases when LE $\to$ 0, then accurate long-term predictions are possible.

In one uncertain division of conditions (complexity vs. chaos) to which rivers can belong, values for KC and LE will be used that have relative meaning. It means that they can have some other but not very different values that were used in this study. Table 5 shows that the streamflow time series of BiH, MB, and the U.S. rivers have a positive LE: (i) BiH [mean 0.250; the range of values (0.186, 0.327)]; (ii) BiH [mean 0.250; the range of values (0.186, 0.327)], MB [mean 0.108; the range of values (0.087, 0.127)]; and (iii) the U.S. [mean 0.111;

range of values (0.011, 0.287)]. In KC and LE domains, we will consider the following values to be "borderlines": $KC \leq 0.520$ (low complex); $KC > 0.520$ (high complex); $LE \leq 0.149$ (low chaotic), and $LE > 0.149$ (high chaotic).

**Table 6.** Mean KC, LE, and LT (in months) values of the monthly streamflow for L, H, and M regimes; (a) eleven gauge stations from seven rivers in Bosnia and Herzegovina, (b) five gauge stations from four rivers in the Mendoza Basin (Argentina), and (c) 1879 gauge stations from the U.S.

(a) Rivers in Bosnia and Herzegovina

| Regime | KC | LE | LT | Mode |
|---|---|---|---|---|
| L | 0.886 | 0.261 | 4.1 | *KL* |
| H | 0.948 | 0.278 | 3.4 | *KL* |
| M | 0.992 | 0.205 | 4.9 | *KL* |

(b) Rivers in the Mendoza Basin

| Regime | KC | LE | LT | Mode |
|---|---|---|---|---|
| M | 0.484 | 0.108 | 9.3 | *kl* |

(c) Rivers in the U.S.

| Regime | KC | LE | LT | Mode |
|---|---|---|---|---|
| L | 0.741 | 0.148 | 6.8 | *Kl* |
| H | 0.696 | 0.142 | 7.0 | *Kl* |
| M | 0.520 | 0.119 | 8.4 | *kl* |

The three-dimensional graph in Fig. 10 clearly visualizes the content of Table 6. There are three distinct clusters of time horizons. The rivers in the Mendoza Basin (lower KC as well as LE) have the highest predictability. Diagonal from this ellipse is an ellipse that encompasses the horizon time of the Bosnia and Herzegovina rivers (high values of KC and LE). The predictability of the U.S. rivers varies from greater predictability (M regime) to somewhat less (L and H regimes). Note that all four modes (*kL*, *Kl*, *kl,* and *Kl*) are covered by this clustering.

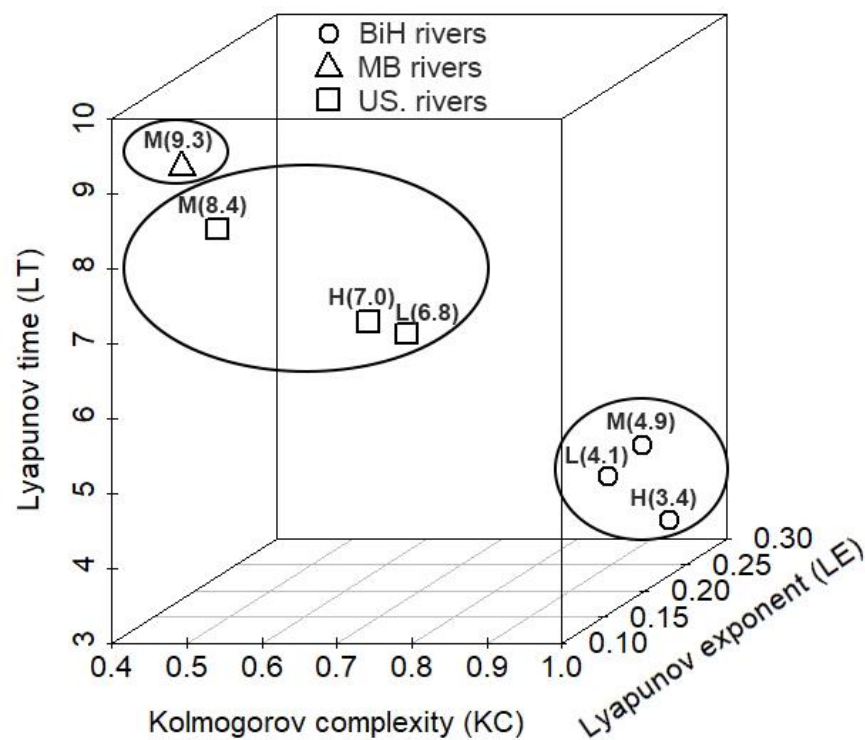

**Fig. 10.** Three-dimensional visualization of prediction horizon (LT) for M, H, and L regimes for rivers in BiH, MB, and the U.S. Numbers in brackets indicate the LT values.

Finally, it should be noted that the actual LE and KC of rivers can vary over time and different sections of a river may exhibit different characteristics. It should also be noted that KC and LE are obtained from long streamflow data bases and that the validity of these measures is limited by the ways in which these measures are calculated. Consequently, if the outcomes are derived from the streamflow at the outlet of a watershed, it may represent the river system at the watershed scale. The only thing that can sometimes be questionable are the criteria for the limit values of KC and LE, which directly determine which mode the streamflow belongs to.

## 5. Conclusions

(1) The relationship between complexity and chaos is an often intriguing but often misunderstood concept. Even though complexity and chaos share some similarities, they are definitely not the same thing. Not all complex systems are chaotic, and not all chaotic systems are inevitably complex. Complex systems are characterized by their intricate organization since the relationships between their components are nontrivial. Complex systems can exhibit chaotic behavior under certain conditions, while chaos can be seen as a subset of complexity, which represents a particular type of complex behavior.

(2) The relationship between chaos and complexity significantly affects the predictability of a system. Chaotic behavior has a tendency to reduce long-term predictability due to the sensitivity to initial conditions. On the other hand, complex systems may exhibit a fusion of predictable and unpredictable behavior that is governed by the presence and magnitude of chaotic elements. Natural fluids, in particular rivers, are complex systems par excellence, which flow in a turbulent regime (Ri in the interval $10^5$-$10^6$) and are highly ranked candidates for exploring the relationship complexity and chaos and its impact on streamflow predictability.

(3) Considering the impact of the complexity and chaotic behavior of the rivers on the streamflow predictability, we realized through the following steps: (i) selection of three regions encompassing a wide range of river types in Bosnia and Herzegovina, the Mendoza

Basin in Argentina and the U.S. (1879 rivers); (ii) creating monthly streamflow time series; (iii) computing the permutation entropy (PE), Kolmogorov complexity (KC), Lyapunov exponent (LE), and Kolmogorov spectra (KC) for all rivers; (iv) making a selection of gauge stations was based on the classification of topography for mountains and other relief classes according to Meybeck et al. (2001): (i) lowlands (0-200 m mean altitude – L type), (ii) platforms and hills (200-500 m – H type), and (iii) mountains with mean elevations between 500 and 6000 m - M type), and calculating averages of all information measures and spectra for all three classification types.

(4) For the considered KC and LE domains of rivers, we used the following values for "borderlines": $KC \leq 0.520$ (low complex - *k*); $KC > 0.520$ (high complex - *K*); $LE \leq 0.149$ (low chaotic - *l*); and $LE > 0.149$ (high chaotic - *L*).

(5) The metrics averaged over all rivers and regimes were: (i) Kolmogorov complexity (KC) - Bosnia and Herzegovina rivers (0.942), the Mendoza Basin rivers (0.484), and the U.S. rivers (0.652); (ii) Permutation entropy (PE) - Bosnia and Herzegovina rivers (0.866), the Mendoza Basin rivers (0.632), and the U.S. rivers (0.836); (iii) (i) Lyapunov exponent (LE) - Bosnia and Herzegovina rivers (0.248), the Mendoza Basin rivers (0.108), and the U.S. rivers (0.136). All rivers have positive LE, i.e., they are in a turbulent regime with a higher level of complexity. A high PE indicates that all rivers exhibit many different ordinal patterns, indicating a high level of complexity, but these patterns can be generated by relatively simple computational processes. A lower number of ordinal patterns in time series typically indicates a higher degree of predictability in the considered system that generated the data. KC is perhaps the most useful measure for estimating the complexity of time series resulting from the evolution of complex systems. A time series can have a large value of PE and lower LE at a time. It means that a high degree of complexity in terms of its ordinal patterns may be explained by a relatively small amount of information or computational complexity. LE, as a measure, indicates the level of chaos in a system.

(6) On the basis of classification (turbulence versus complexity), as given in (4), we derived the following time horizons for the considered rivers: (i) in Bosnia and Herzegovina [all rivers are in the *KL* mode with values of LT (in months): 4.9 (M), 3.4 (H) and 4.1 (L)]; (ii) in the Mendoza Basin [with values of LT (in months): 9.3 (M, the *kl* mode); and in the U.S. [with values of LT (in months): 8.4 (M, *kl*), 7.0 (H, *Kl*) and 6.8 (L, *Kl*)].

**Author contributions**

**Dragutin T. Mihailović**: Conceptualization, Formal analysis, Investigation; Methodology; Project administration, Resources, Software, Supervision; Validation, Writing - original draft; Writing - review & editing **Slavica Malinović-Milićević**: Formal analysis, Investigation, Methodology, Software, Validation, Visualization, Writing - review & editing **Francisco Javier Frau:** Data curation, Investigation, Formal analysis, Writing - original draft; Writing - review & editing **Jeongwoo Han**: Data curation, Investigation, Formal analysis, Writing - original draft; Writing - review & editing **Vijay P. Singh**: Conceptualization, Funding acquisition, Formal analysis, Supervision, Validation, Writing - review & editing

**Declaration of Competing Interest**

The authors declare that they have no known competing financial interests or personal relationships that could have appeared to influence the work reported in this paper.

**Data availability**

The data authors used are publicly available online: Monthly naturalized streamflow data (at https://www.sciencebase.gov/catalog/item/59cbbd61e4b017cf314244e1), NOAA nClimGrid monthly precipitation and temperature data (at https://www.ncei.noaa.gov/access/metadata/landing-page/bin/iso?id=gov.noaa.ncdc:C00332), National Inventory Dams data (at https://www.fema.gov/emergency-managers/risk-management/dam-safety/national-inventory-dams), The mean slope of watershed from EPA (at https://www.epa.gov/wsio/wsio-indicator-data-library), Elevation of gauge station from USGS (at https://waterdata.usgs.gov/nwis/sw).